\documentclass[a4paper,aps,manuscript]{revtex}
\tightenlines
\usepackage[dvips]{graphics}
\begin{document}
\draft

\title{An Efficient Modified \lq\lq Walk On Spheres" Algorithm for the Linearized Poisson-Boltzmann Equation}

\author{Chi-Ok Hwang and Michael Mascagni}

\address{Department of Computer Science, Florida State University, 203 Love Building Tallahassee, FL 32306-4530}

\date{\today}
\maketitle
\begin{abstract}
A discrete random walk method on grids was proposed and used to solve the linearized 
Poisson-Boltzmann equation (LPBE)~\cite{Rammile}. Here, we present a new and efficient grid-free
 random walk method.
Based on a modified \lq\lq Walk On Spheres" (WOS) algorithm~\cite{Elepov-Mihailov1973} for the LPBE, 
this Monte Carlo algorithm uses a survival 
probability distribution function for the random walker in a continuous and free diffusion region. 
The new simulation method is illustrated by computing
four analytically solvable problems. In all cases, excellent agreement is observed.
\end{abstract}

\newpage
Random walk methods have been used to solve a wide variety of parabolic and elliptic 
partial differential equations (PDEs)~\cite{Rammile,Muller56,ZC89,Sabelfeld91,TK92}.
Generally, there are two broad classes of random walk methods; one uses 
discrete random walks on grids~\cite{Rammile}, and 
the other continuous random walks in free space~\cite{Muller56,ZC89,Sabelfeld91,TK92}.
One of the widely used continuous random walk methods, the \lq\lq Walk On Spheres" (WOS) 
method~\cite{Muller56,Booth82,Booth,Mikhailov1995}, uses the first-passage 
probability distribution on a sphere to facilitate large steps in random walks.
(The first-passage probability, $w({\bf x};{\bf x_0})$, is the probability
of hitting the vicinity of ${\bf x}$ on the bounding surface for the
first time when the random walker starts from ${\bf x_0}$, a point inside the bounding surface.)
This continuous random walk method needs to discretize neither space nor time, nor the diffusing 
trajectory, and so it is 
particularly advantageous when the geometry of the region of interest is very complex or 
if the solution of the PDE is required at only a relatively small number of points.

We are interested in solutions to the Dirichlet problem for the linearized Poisson-Boltzmann
equation (LPBE) in the domain $\Omega$:

\begin{equation}
\nabla^{2}\psi({\bf x}) = \kappa^2\psi({\bf x}),  \qquad      {\bf x} \in \Omega,
\end{equation}
\begin{equation}
\psi({\bf x}) = \psi_0({\bf x}),          \qquad          {\bf x} \in \partial\Omega,
\end{equation}
\noindent
where $\kappa$ is called the inverse Debye length~\cite{Rammile}.
Notice that when $\kappa^2$ is zero, the above problem becomes a Dirichlet problem for the
Laplace equation. The WOS method for the Dirichlet problem for the Laplace equation has been widely 
used~\cite{Muller56,Booth82,Booth,ZC89,TK89,Kim-Torquato1991}. We will combine this
WOS method with a survival probability density function which incorporates the term
involving $\kappa^2$ in the LPBE.

In a discrete random walk method~\cite{Rammile} for the LPBE, the corresponding Master equation 
relates $\kappa^2$ to the removal
probability of the random walker on the grid. During each step of the discrete random walk,
the walker either moves to one of the neighboring sites, or stays fixed, or is removed.
This probabilistic interpretation of
$\kappa^2$ can be also extended to continuous random walk methods once we know the
survival probability distribution function of a random walker in continuous space.

\begin{figure}[t]
\vspace{0.3cm}
{\centering \includegraphics{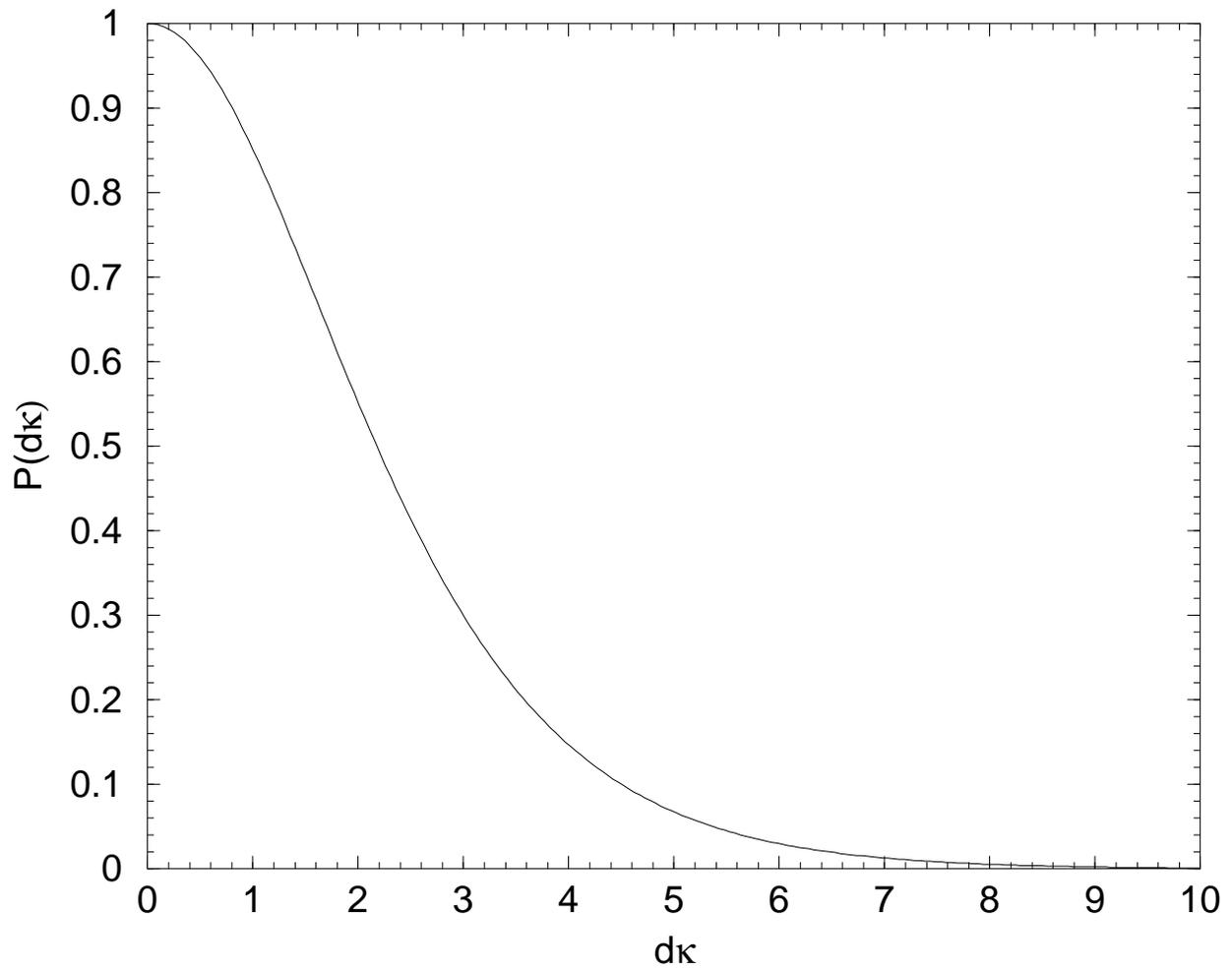} \par}
\vspace{0.3cm}
\caption{The survival probability density function; $d$ 
is the diffused distance of a random walker from the starting position,
and $\kappa$ is the inverse Debye length.
}
\label{removing}
\end{figure}

In this letter, we obtain this survival probability distribution function of a random walker 
in continuous space by reinterpreting the weighting function in the previous modified WOS
method~\cite{Elepov-Mihailov1973} for the LPBE (see the Appendix for more details). 
The survival probability of a random walker in a continuous and free diffusion region is
given by~\cite{Elepov-Mihailov1973}:

\begin{equation}
p(d) = d{\kappa}/\sinh(d{\kappa}),
\end{equation}
\noindent 
where $d$ is the distance from the starting point in the diffusion region.  
Figure~\ref{removing} shows this probability density function.
We modify the WOS method to incorporate the survival probability to solve the
LPBE via a continuous random walk method.
This probability density combined with the WOS method is used to remove a random walker 
during the random walk
by the acceptance-rejection method~\cite{Hammersley-Handscomb64}. We generate
a random number, $\eta$ in $[0,1)$ when we perform a WOS step, and we compare $\eta$
 with $p(d)$, the survival probability at $d$, the radius of WOS. If $\eta > p(d)$, 
the random walker is removed at this WOS step.

An estimate for the solution of the LPBE at ${\bf x_0}$, 
where random walkers start, is given by $S_{N}$:

\begin{equation}
S_{N} = \frac{1}{N} \sum_{i=1}^{N_s} \psi_0(X_{n_i}).
\end{equation}

\noindent
Here, $N$ is the total number of random walkers, $N_s$ is the number of survived-and-absorbed 
random walkers, and $X_{n_i}$ is the final position of the walker on the
boundary when it is absorbed after $n_i$ WOS steps.

In this method, like the WOS method, errors are due to both statistical sampling
and the $\delta$-absorption layer which captures random walkers near the boundary to terminate
their random walk. However, the error from the $\delta$-absorption layer
can always be made smaller than the statistical error~\cite{Booth82,Booth}.
For the same random walk, the estimate difference between using $\delta$ and
$\delta/10$ gives a measure of the error due to the finite width of the $\delta$-absorption layer.
By adjusting $\delta$
we can make the error from the absorption layer less than the statistical error.
This means that if we increase the number of random walkers to decrease the statistical error, 
consequently we must reduce $\delta$ and so increase the running time.

\begin{figure}[t]
\vspace{0.3cm}
{\centering \includegraphics{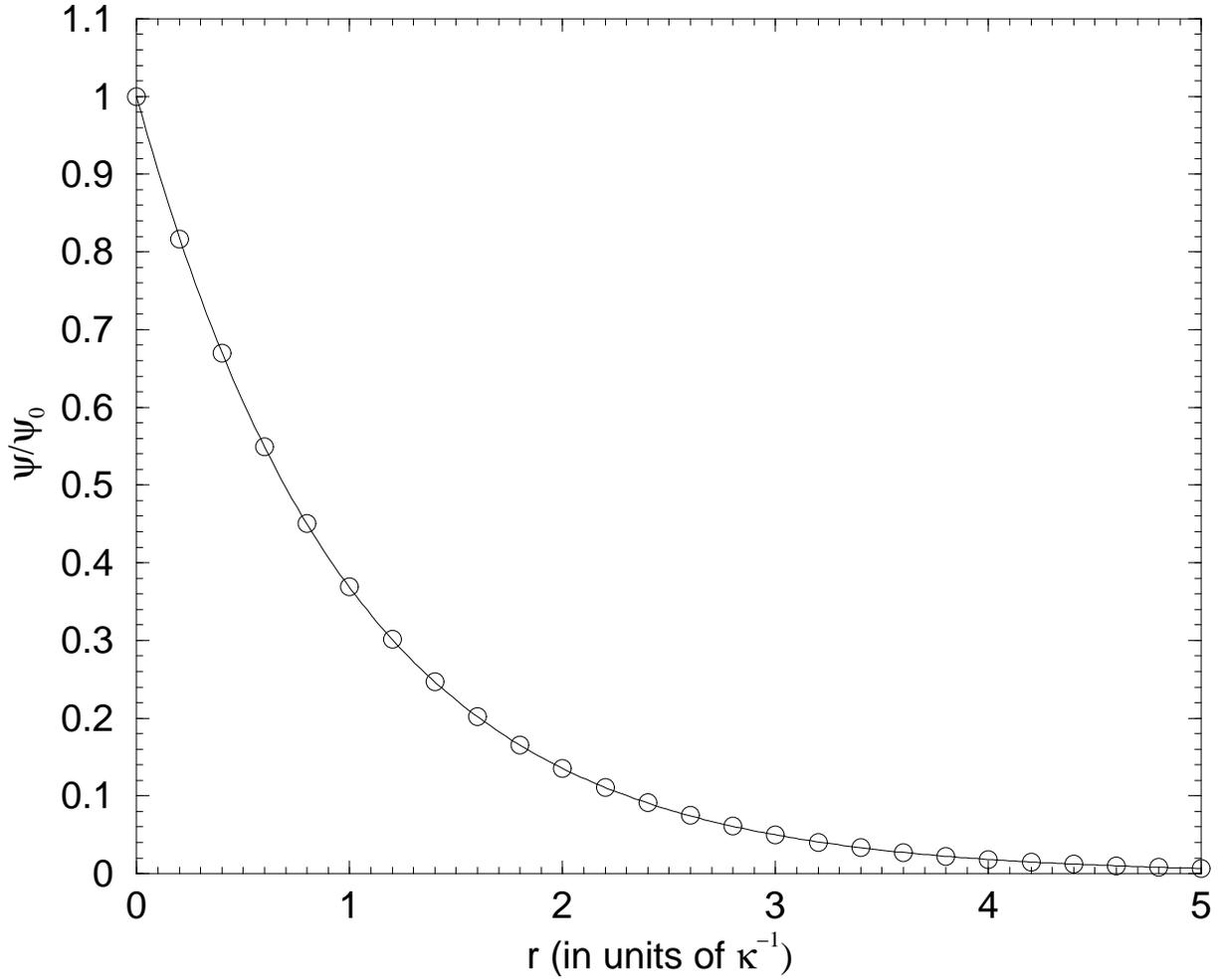} \par}
\vspace{0.3cm}
\caption{The electric potential away from an charged infinite flat plate in an 0;
the solid line is the analytic solution and the circles are the simulation
results with $10^5$ random walks and the absorption layer $\delta=10^{-4}$. 
Here, $r$ is the distance to the plate and $\kappa$ the inverse Debye length.
}
\label{plate}
\end{figure}

\begin{figure}[t]
\vspace{0.3cm}
{\centering \includegraphics{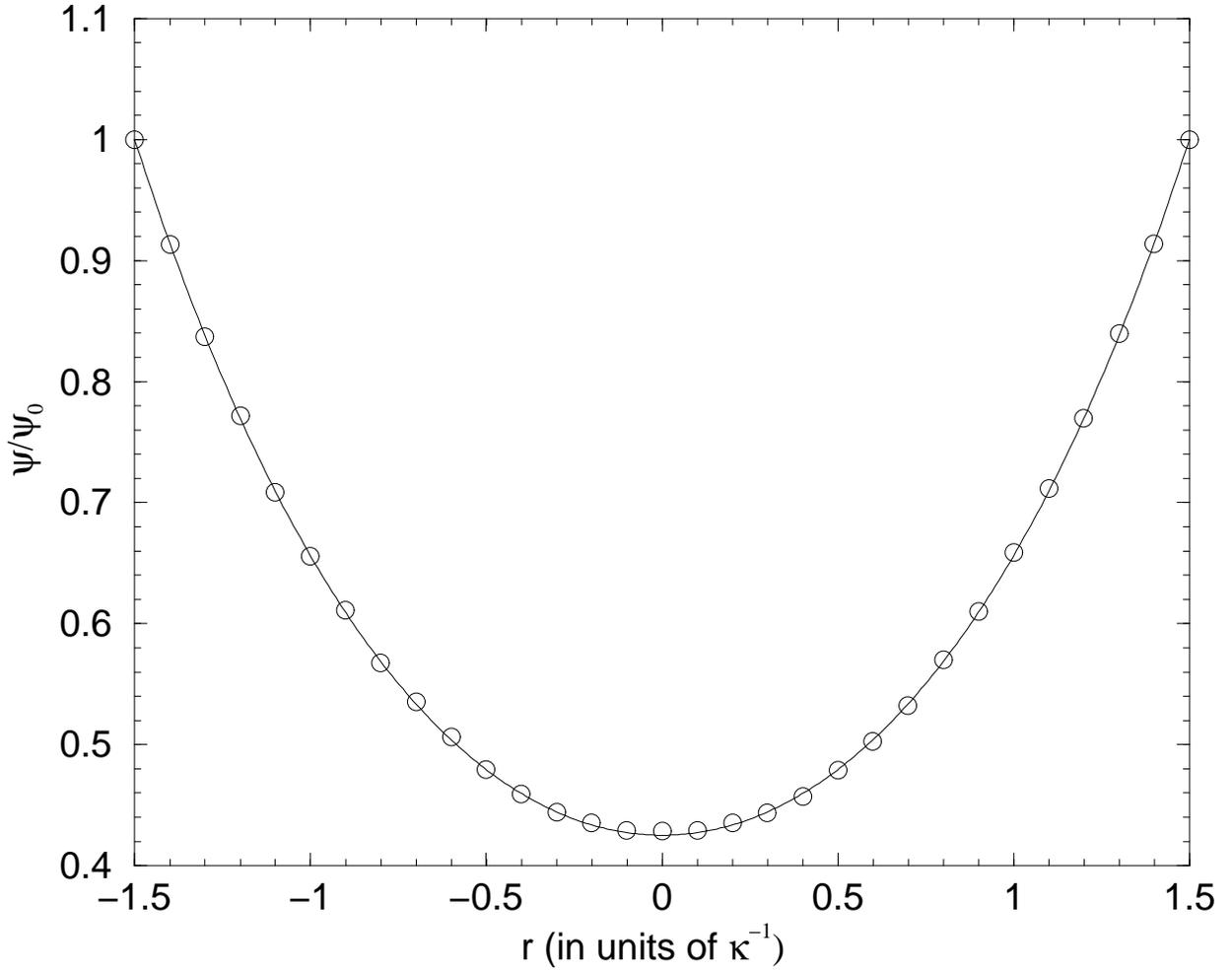} \par}
\vspace{0.3cm}
\caption{The electric potential in an electrolyte between two infinite charged parallel flat plates;
the solid line is the analytic solution and the circles are the simulation
results with $10^5$ random walks and the absorption layer $\delta=10^{-4}$. 
Here, $r$ is the distance from the mid-point of the plates and 
$\kappa$ the inverse Debye length.
}
\label{plates}
\end{figure}

\begin{figure}[t]
\vspace{0.3cm}
{\centering \includegraphics{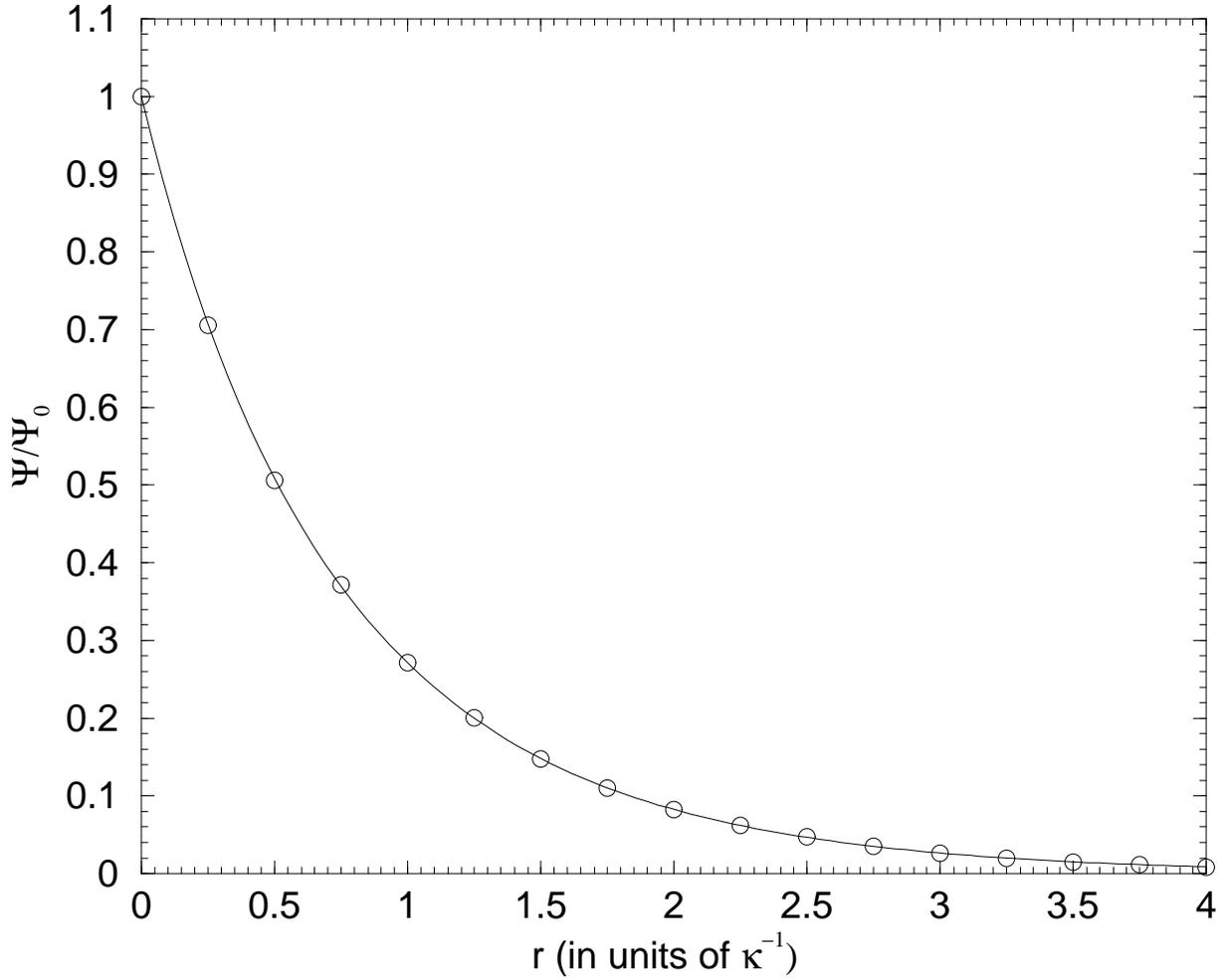} \par}
\vspace{0.3cm}
\caption{The electric potential away from an infinitely long charged cylinder in an electrolyte;
the solid line is the analytic solution and the circles are the simulation
results with $10^5$ random walks and the absorption layer $\delta=10^{-4}$. 
Here, $r$ is the distance from the surface of the cylinder with unit radius
 and $\kappa$ the inverse Debye length.
}
\label{cylinder}
\end{figure}

\begin{figure}[t]
\vspace{0.3cm}
{\centering \includegraphics{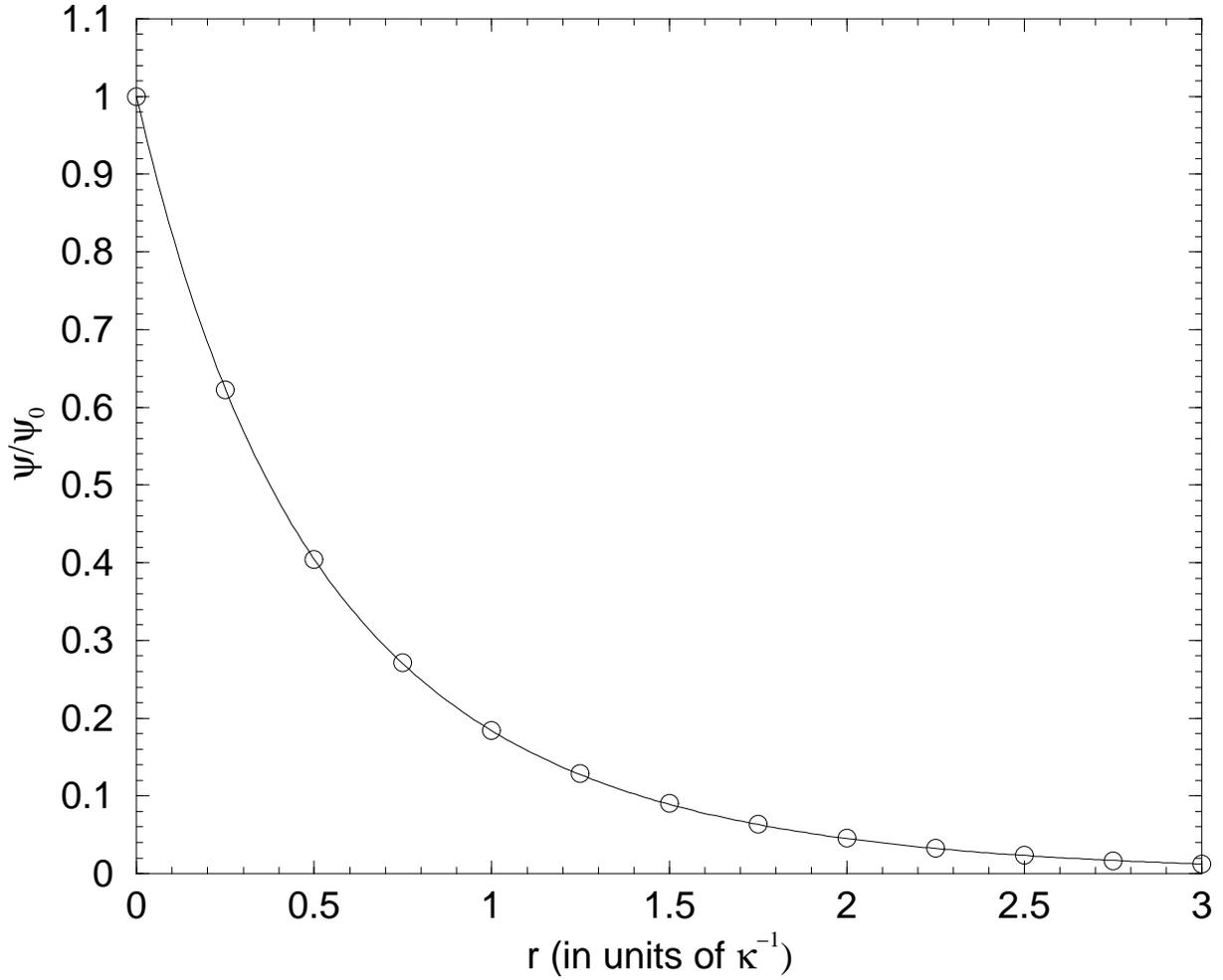} \par}
\vspace{0.3cm}
\caption{The electric potential away from the surface of a charged sphere in an electrolyte;
the solid line is the analytic solution and the circles are the simulation
results with $10^5$ random walks and the absorption layer $\delta=10^{-4}$. 
Here, $r$ is the distance from the surface of the sphere with unit radius
 and $\kappa$ the inverse Debye length.
}
\label{sphere}
\end{figure}

In the following, we compare our simulation results with the analytic results
for four problems, which were used as examples for the discrete random walk method~\cite{Rammile}. 
In all cases, the results are given as those normalized by the boundary condition
 $\psi_0$, which is assumed sufficiently small for the LPBE to be valid.
The number of random walks used for the solution at a point is $10^5$, and the absorption layer
thickness is $\delta= 10^{-4}$.  The analytic results~\cite{Rammile} are shown with solid lines
in Figs.~\ref{plate},~\ref{plates},~\ref{cylinder} and~\ref{sphere} and  
our simulation results with circles.
For the all four cases, our simulation results show excellent agreement.
%
%
%
%
%
%
%
%
%

Our method has several features. First, it is easier to implement and 
will be faster than the other discretized methods, such as 
the discrete random walk method~\cite{Rammile}, the finite difference 
method~\cite{Nicholls-Honig1991} and the boundary-element method~\cite{Yoon-Lenhoff1990},
especially with complicated geometries.
For a desired point, it takes only a few seconds to compute a solution 
with $10^5$ random walkers and $\delta =10^{-4}$ 
on a 550 Mhz PC.  However, it is hard to compare to other methods 
because they compute  solutions at all grid points. We can safely say that continuous Monte Carlo
methods are more efficient when the solution is required only at relatively small number of points. 
Secondly, the accuracy and the running time of our method depends primarily on the number of 
statistical samples, and so it is naturally parallel. 
Thirdly, it is certain that our new method is faster than the old modified WOS 
method~\cite{Elepov-Mihailov1973}, because while some of our random walkers are removed
during their random walk,
in the old method all random walkers must complete their random walks to contribute
to the solution according to their weightings. Also, in open boundary cases, like the three examples
except the parallel plates, it is necessary to use a certain cut-off 
in the old modified WOS~\cite{Elepov-Mihailov1973} to kill random walkers, 
which will bias the results. 
As an example, in Table~1 in the case of the parallel plates, we compare our new method with
the old modified WOS method,~\cite{Elepov-Mihailov1973}. We use the customary comparison
method for Monte Carlo methods, the time consumption (or laboriousness)~\cite{Sobol1994}: 
$t\times {\b D}\xi$, 
where $t$ is the CPU time expended in calculating a single estimate and ${\b D}\xi$ is
the variance of the estimates. 
The less laborious the algorithm, the more efficient it is.
In Table~1, the time consumption (or laboriousness) of our algorithm is better 
than that of the old modified WOS method.
Finally, our method is easy to extend to solve the LPBE with source 
terms~\cite{Elepov-Mihailov1973}. 
That will be the subject of our upcoming research with biochemical applications.

{\centering \textbf{APPENDIX} \par}
In this appendix, we show how the weighting function in the old modified WOS 
method~\cite{Elepov-Mihailov1973} can be
interpreted as the survival probability distribution function.
For simplicity, consider the LPBE  in the old modified WOS method~\cite{Elepov-Mihailov1973}.
The solution at ${\bf x_0}$ in the domain can be expressed 
as follows~\cite{Elepov-Mihailov1973}: 

\begin{equation}
u({\bf x_0}) = \frac{1}{N}\sum_{i=1}^{N}Q_i^{n_i}\psi_0(X_{n_i}),
\end{equation}
\noindent
where
\begin{equation}
Q_i^0 = 1, \qquad Q_i^{n_i} = Q_i^{n_i-1}\frac{d_i^{n_i-1}\kappa}{\sinh(d_i^{n_i-1}\kappa)}, \qquad d_i^{n_i}=d(P_i^{n_i}). 
\end{equation}

\noindent
Here, $N$ is the total number of diffusing random walkers, $i$ refers to $i$th random walker, 
 $X_{n_i}$ is the position where the $i$th random walker is absorbed in the $\delta$-absorption
layer after $n_i$ WOS steps, and $d_i^{n_i}$ the radius of $n_i$th WOS
of the $i$th random walker. 

If we interpret $Q_i^{n_i}$ as a survival probability of $i$th random walker, 
$\sum_{i=1}^{N}Q_i^{n_i}$ is the total number of survived-and-absorbed random walkers.
Furthermore, due to the property of probabilistic random sampling from
the total random walkers, only the survived-and-absorbed random walkers can be regarded 
as contributors to the solution.
This reinterpretation of the weighting function as the survival probability distribution function
is a kind of the fractional sampling method, 
{\it i.e.} \lq Russian Roullete\rq,~\cite{Hammersley-Handscomb64}
 which has been used extensively in neutron transport and similar problems.


\begin{thebibliography}{10}

\bibitem{Rammile}
R.~Ettelaie.
\newblock Solutions of the linearized poission-boltzmann equation through the
  use of random walk simulation method.
\newblock {\em Journal of Chemical Physics}, 103(9):3657--3667, 1995.

\bibitem{Muller56}
Mervin~E. M{\"u}ller.
\newblock Some continuous {M}onte {C}arlo methods for the {D}irichlet problem.
\newblock {\em Annals of Mathematical Statistics}, 27:569--589, 1956.

\bibitem{ZC89}
L.~H. Zheng and Y.~C. Chiew.
\newblock Computer simulation of diffusion-controlled reactions in dispersions
  of spherical sinks.
\newblock {\em Journal of Chemical Physics}, 90(1):322--327, 1989.

\bibitem{Sabelfeld91}
K.~K. Sabelfeld.
\newblock {\em Monte Carlo Methods in Boundary Value Problems}.
\newblock Springer-Verlag, Berlin, 1991.

\bibitem{TK92}
S.~Torquato and I.~C. Kim.
\newblock Cross-property relations for momentum and diffusional transport in
  porous media.
\newblock {\em Journal of Applied Physics}, 72(2):2612--2619, 1992.

\bibitem{Booth82}
T.~E. Booth.
\newblock Regional {M}onte {C}arlo solution of elliptic partial differential
  equations.
\newblock {\em Journal of Computational Physics}, 47:281--290, 1982.

\bibitem{Booth}
T.~E. Booth.
\newblock Exact {M}onte {C}arlo solution of elliptic partial differential
  equations.
\newblock {\em Journal of Computational Physics}, 39:396--404, 1981.

\bibitem{Mikhailov1995}
G.~A. Mikhailov.
\newblock {\em New {M}onte {C}arlo Methods with Estimating Derivatives}.
\newblock VSP, Utrecht, Netherlands, 1995.

\bibitem{TK89}
S.~Torquato and I.~C. Kim.
\newblock Efficient simulation technique to compute effective properties of
  hetergeneous media.
\newblock {\em Applied Physics Letters}, 55:1847--1849, 1989.

\bibitem{Kim-Torquato1991}
I.~C. Kim and S.~Torquato.
\newblock Effective conductivity of suspensions of hard spheres by {B}rownian
  motion simulation.
\newblock {\em Journal of Applied Physics}, 69(4):2280--2289, 1991.

\bibitem{Elepov-Mihailov1973}
B.~S. Elepov and G.~A. Mihailov.
\newblock The \lq\lq {W}alk {O}n {S}pheres" algorithm for the equation $\Delta
  u -cu= -g$.
\newblock {\em Soviet Math. Dokl.}, 14:1276--1280, 1973.

\bibitem{Hammersley-Handscomb64}
J.~M. Hammersley and D.~C. Handscomb.
\newblock {\em {M}onte {C}arlo methods}.
\newblock Methuen \& Co. Ltd., London, 1964.

\bibitem{Nicholls-Honig1991}
A.~Nicholls and B.~Honig.
\newblock A rapid finite difference algorithm, utilizing successive
  over-relaxation to solve the poisson-boltzmann equation.
\newblock {\em Journal of Computational Chemistry}, 12:435--445, 1991.

\bibitem{Yoon-Lenhoff1990}
B.~J. Yoon and A.~M. Lenhoff.
\newblock A boundary element method for molecule electrostatics with
  electrolyte effects.
\newblock {\em Journal of Computational Chemistry}, 11:1080--1086, 1990.

\bibitem{Sobol1994}
I.~M. Sobol.
\newblock {\em A Primer for the {M}onte {C}arlo Method}.
\newblock CRC Press, Washington, USA, 1994.

\end{thebibliography}

\newpage
\begin{table}[t]
\caption{Time consumption comparison of our algorithm with the old
modified WOS in the case of parallel plates at the mid-point; 
the variances are obtained from 100 independent runs, 
the number of random walks per run is $10^5$ and the absorption layer, $\delta = 10^{-4}$.
}
\begin{tabular}{|c|c|c|c|}
\hline
method           & CPU time per run (secs) &   variance        &  time consumption      \\ \hline
old method  & $13.47$                   & $4.63\times 10^{-7}$  &  $6.24\times 10^{-6}$  \\
new method                 &  $2.97$     & $1.98\times 10^{-6}$  &  $5.88\times 10^{-6}$  \\
\hline
\end{tabular}
\end{table}

\end{document}